\documentclass[sigplan]{acmart}
\usepackage{pgfplots}
\usepackage{makecell}
\usepackage{mathtools}
\usepackage{listings}
\usepackage{algorithmic}
\usepackage[boxed]{algorithm}
\usepackage{wrapfig}    
\usepackage{multirow}

\usepackage[colorinlistoftodos]{todonotes} 

\newcommand{\Lyngso}{Lyngs\o}
\newcommand{\expnumber}[2]{{#1}\mathrm{e}{#2}}

\definecolor{darkgreen}{RGB}{0,153,0}

\lstdefinestyle{CStyle}{
    basicstyle=\footnotesize,
    keywordstyle=\bfseries,
    morekeywords={assert, for, if, is, to, S0, S1, S2, S3, S4, C0, C1, C2, C3, C4, C5, inputs, outputs, locals, let, case, and, or, affine},
    moredelim=**[is][\color{red}]{@r@}{@r@},
    moredelim=**[is][\color{blue}]{@b@}{@b@},
    moredelim=**[is][\color{green}]{@g@}{@g@},
    moredelim=**[is][\color{gray}]{@k@}{@k@},
    frame=single
}

\AtBeginDocument{%
  }

\setcopyright{acmlicensed}
\copyrightyear{2018}
\acmYear{2018}
\acmDOI{XXXXXXX.XXXXXXX}

\pgfplotsset{compat=1.18}
\begin{document}

\title{Simplification of Polyhedral Reductions in Practice}

\author{Louis Narmour}
\email{louis.narmour@cololstate.edu}
\orcid{0009-0009-3298-5282}
\affiliation{
  \institution{Colorado State University}
  \city{Fort Collins}
  \state{Colorado}
  \country{USA}
}
\affiliation{
  \institution{University of Rennes, Inria, CNRS, IRISA}
  \city{Rennes}
  \country{France}
}

\author{Ryan Job}
\email{ryan.job@colostate.edu}
\orcid{0009-0002-4581-3737}
\affiliation{%
  \institution{Colorado State University}
  \city{Fort Collins}
  \state{Colorado}
  \country{USA}
}

\author{Tomofumi Yuki}
\affiliation{%
 \institution{Unaffiliated}
 \country{Japan}
}

\author{Sanjay Rajopadhye}
\email{sanjay.rajopadhye@colostate.edu}
\orcid{0000-0002-4246-6066}
\affiliation{%
  \institution{Colorado State University}
  \city{Fort Collins}
  \state{Colorado}
  \country{USA}
}

\renewcommand{\shortauthors}{Narmour et al.}

\begin{abstract}
Reductions combine collections of inputs with an associative (and here, also commutative) operator to produce collections of outputs.  When the same \emph{value} contributes to multiple outputs, there is an opportunity to reuse partial results, enabling \emph{reduction simplification}.  We provide the first complete push-button implementation of reduction simplification in a compiler.  We evaluate its effectiveness on a range of real-world applications, 
and show that simplification rediscovers several key results in algorithmic improvement across multiple domains, previously only obtained through clever manual human analysis and effort.  We also discover alternate, previously unknown algorithms, albeit without improving the asymptotic complexity.
\end{abstract}




\maketitle

\section{Introduction}

Computing technology has become increasingly powerful and complex over the years, offering more capability with each new generation of processors.
The latest generation of Intel Xeon processors, for example, supports configurations up to 50 cores on a single die with 100 MB of last-level cache.\footnote{\url{https://www.intel.com/content/www/us/en/products/sku/231750/intel-xeon-platinum-8468h-processor-105m-cache-2-10-ghz/specifications.html}}
However, using all the available processing power in the presence of complex data dependencies is not always easy.
Relying on traditional compilers to produce high-performance code for a given input program is insufficient.
One reason is that general-purpose language compilers must be very conservative in the types of optimizations that can be employed.
Consequently, the onus is on the application developer to write the program in such a way that the compiler can successfully detect optimization opportunities. 
This is challenging because it is often easier to think about problems from a higher level of abstraction, whereas writing efficient code requires lower-level reasoning.
Over the years, this has led to  a wide variety of Domain Specific Languages (DSLs) and highly specialized frameworks.

In this paper, we study the optimization of programs that can be specified by reductions within the polyhedral model~\cite{rajopadhye_synthesizing_1989, feautrier_dataflow_1991, feautrier_efficient_1992, amarasinghe_communication_1993, fortes_data_1984, irigoin_supernode_1988, lam_systolic_1989, lengauer_loop_1993, pugh_omega_1991, quinton_mapping_1989, ramanujam_nested_1990, schreiber_automatic_1990, wolf_loop_1991, wolf_data_1991, wolfe_iteration_1987}.
Reductions are ubiquitous in computing and typically involve applying an associative, often commutative, operator to collections of inputs to produce collections of results.
Such operations are interesting because they often require special handling to obtain good performance.
The OpenMP C/C++ multithreading API~\cite{openmp_architecture_review_board_openmp_2021} even has directives tailored specifically for parallelizing 
reductions.
Typical compiler optimizations yield, at best, a constant fold speedup.  However, in some cases, the input program specification may involve \textit{reuse}---the same value contributing to multiple results.  When properly exploited, it is possible to improve the asymptotic complexity of such programs.  A complexity improvement from, say, $O(N^3)$ to $O(N^2)$ yields unbounded speedup since, asymptotically, $N$ can be arbitrarily large.  Optimizations of this type have traditionally relied on very clever and manual human analysis and engineering effort.

Gautam and Rajopadhye~\cite{gautam_simplifying_2006} previously showed how to reduce, by such polynomial degrees, the asymptotic complexity of polyhedral reductions.  They developed a transformation called \emph{simplification} and outlined, without providing details, a recursive algorithm to automatically simplify polyhedral reductions.  Their work (henceforth referred to as GR06) is purely theoretical.
We address these issues, and in doing so, we make the following contributions:

\begin{itemize}
\item Our implementation includes a heuristic algorithm for choosing the specific reuse to exploit at each step of the recursion, thereby realizing a complete push-button simplification engine.
\item Using the integer set library (\texttt{isl})~\cite{verdoolaege_isl_2010}, we provide an open source implementation to construct the \emph{face lattice} (see Section~\ref{sec:background-face-lattice}) of arbitrary parameterized convex polytopes.  This is a critical data structure for simplification but has other uses beyond that, notably in program verification~\cite{allamigeon_formalizing_2022, boulme_verified_2018}.
\end{itemize}


We evaluate the effectiveness of our implementation on a range of programs involving reductions and provide an accompanying software artifact as a proof of concept.
Interestingly, our simplification of an $O(N^{4})$ RNA secondary structure prediction algorithm produces \textit{four} distinct versions, all with cubic complexity. 
One of these rediscovers the 1999 result of \Lyngso{} et al.  (called ``fast-i-loops'' algorithm), while the other three were previously unknown.

The remainder of the paper is organized as follows.
In Section~\ref{sec:motivating-examples}, we provide examples that motivate our work.
Then, in Section~\ref{sec:background}, we review the relevant background on simplification and state the open problems.  In Section~\ref{sec:implementation}, we review how we address these problems and explain the details of the implementation of our compiler.  We provide an analytical and quantitative evaluation in Sections~\ref{sec:case-studies} and \ref{sec:eval-rna}. Sections~\ref{sec:related-work} and~\ref{sec:future-work} discuss related work and open problems. Finally, we conclude in Section~\ref{sec:conclusion}.

\section{Motivating Examples} \label{sec:motivating-examples}

We now summarize two applications involving reductions with reuse, which were previously optimized manually.  We emphasize that these are non-trivial examples. The first is taken from a classic paper on Algorithm-Based Fault Tolerance (ABFT), and the second is the fast-i-loops algorithm.

\subsection{Fault Tolerant Matrix Multiplication} \label{sec:motivating-examples-abft}

Algorithm-based fault tolerance (ABFT) was first proposed in 1984 by Huang and Abraham~\cite{huang_algorithm-based_1984} and provides an elegant way to detect and correct silent data corruption errors.
The main idea is to augment the computation with extra work using invariant checksums (i.e., reductions) by exploiting algebraic identities, which should remain constant-valued.
By comparing checksums of the result with their separately computed predicted values, errors can be flagged if their difference rises above some threshold.
However, doing this in practice requires effort and clever analysis of the particular target application.
Historically, this has led to a completely separate work for each ``new'' application of ABFT: convolutional neural networks~\cite{zhao_ft-cnn_2021}, LU factorization~\cite{davies_correcting_2018}, Cholesky decomposition~\cite{hakkarinen_fail-stop_2015}, and dense linear algebra~\cite{wu_towards_2016}, just to name a few.  Our simplifying compiler can automatically generate fault-tolerant codes.

Consider the classic matrix multiplication for square $N$-by-$N$ matrices where $C = A \times B$, specified by the equation:
\begin{equation} \label{eq:mm}
    C_{i,j} = \sum_{k=1}^{N} \big( A_{i,k} \times B_{k,j} \big)
\end{equation}

\begin{figure}[tbh]
    \centering
    \includegraphics[width=0.47\textwidth]{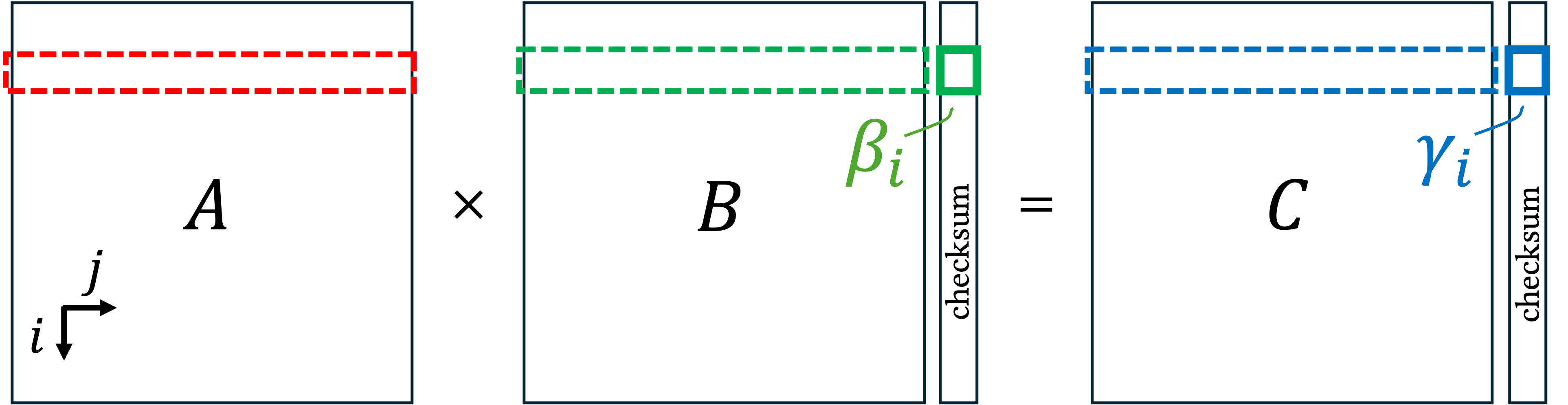}
    \caption{ABFT checksums, (from Huang and Abraham~\cite{huang_algorithm-based_1984}).}
    \Description{ABFT checksums).}
    \label{fig:abft-mm}
\end{figure}

Huang and Abraham defined $\beta_{i}$, $\gamma_{i}$ and $\gamma'_{i}$ as, respectively, the row checksums of $B$ and $C$, and the inner product of a row of $A$ and the checksum vector $\beta$ as defined in Equations \ref{eq:abft-gamma} and \ref{eq:abft-gamma-prime} (and illustrated in Figure~\ref{fig:abft-mm}).
\begin{equation} \label{eq:abft-gamma}
    \textcolor{blue}{\gamma_{i}} = \sum_{j=1}^{N} \textcolor{blue}{C_{i,j}}
\end{equation}
\begin{equation}\label{eq:abft-gamma-prime}
    \gamma_{i}' = \sum_{j=1}^{N} \big( \textcolor{red}{A_{i,j}} \times \beta_{j} \big), \hspace{8mm} \textcolor{darkgreen}{\beta_{i}} = \sum_{j=1}^{N} \textcolor{darkgreen}{B_{i,j}}
\end{equation}

The two quantities $\gamma_{i}$ and $\gamma'_{i}$ compute the same numerical value, and errors occurring during the computation of Equation~\ref{eq:mm} can be detected when their difference, $|\gamma_{i}-\gamma_{i}'|$, is sufficiently large.
More importantly, the $O(N^{2})$ cost of computing Equations~\ref{eq:abft-gamma} and~\ref{eq:abft-gamma-prime} is \textit{cheap} relative to the main $O(N^{3})$ computation in Equation~\ref{eq:mm}.

The complete derivation of Equation~\ref{eq:abft-gamma-prime}, discoverable by our compiler, is given in Section~\ref{sec:abft-derivation}.
The key insight here is that the task of constructing ABFT checksums can be cast as an instance of the simplifying reductions problem.
Consequentially, cheap checksum specifications can be obtained systematically from simplification.

\subsection{RNA Secondary Structure: Fast-i-Loops} \label{sec:motivating-examples-1999-lyngso}

Ribonucleic acid (RNA) forms one of the building blocks of life.  It is described as a sequence of bases drawn from a 4-letter alphabet.  An important determiner of the molecule's function is its \emph{secondary structure}, which indicates the bases that are bonded together.
One method for secondary structure prediction uses thermodynamic equations to determine the minimum free energy configuration (i.e., the most likely to occur in nature).  These equations are defined as a system of dynamic programming recurrence equations.  The original definition~\cite{zuker_optimal_1981} populates these tables with a time complexity of $O(N^3)$, with one notable exception that calculates the energy of an \textit{internal loop} structure in $O(N^4)$ time. Equation~\ref{eq:rna-vbi-n4} is a representative example of the same form as this equation. It computes a 2-dimensional result, $Y_{i,j}$, as the minimum of all pairs of points between $i$ and $j$.
\begin{equation}\label{eq:rna-vbi-n4}
    Y_{i,j} =
    \min_{i<p<q<j}\bigl(
        A_{p,q} +
        B_{p-i+j-q} +
        C_{|p-i-j+q|}
    \bigr)
\end{equation}

In 1999, \Lyngso{} et al.~\cite{lyngso_fast_1999} exploited an invariant property in Equation~\ref{eq:rna-vbi-n4} to rewrite it essentially as Equations~\ref{eq:rna-vbi-n3} and~\ref{eq:rna-vbiprime}, which have $O(N^3)$ complexity.  A derivation of how our simplifier produces this \textit{fast-i-loops} algorithm is in Section \ref{sec:eval-rna}.

\begin{gather}
    Y_{i,j} = \min_{1 < k < j-i} \bigl( B_{k} + Z_{i,j,k} \bigr)
    \label{eq:rna-vbi-n3}
    \\
    Z_{i,j,k} = \min \begin{pmatrix}
        A_{i+1, j-k+1} + C_{|-k+2|} \\
        A_{i+k-1,j-1} + C_{|k-2|} \\
        Z_{i+1,j-1,k-2}
    \end{pmatrix}
    \label{eq:rna-vbiprime}
\end{gather}

Since its discovery, there have been several implementations of fast-i-loops~\cite{dirks_partition_2003,fornace_unified_2020,jacob_rapid_2010}, each requiring painstaking effort by the authors to first transform the specification into a more usable format and then write the code that implements the algorithm.  Our compiler takes the $O(N^4)$ specification of Equation \ref{eq:rna-vbi-n4} and automatically discovers four $O(N^3)$ improved algorithms.

\section{Background} \label{sec:background}

We first summarize the GR06 simplification algorithm with an example and then highlight the challenges that arise during implementation.  We use the following terminology:
\begin{itemize}
    \item \textit{Polyhedron}: A set of integer points defined by a list of inequality and equality constraints.
    \item \textit{Reduction body} ($\mathcal{D}$): A $d$-dimensional polyhedron representing the values of the program variable indices.
    \item \textit{Facet}: A $k$-dimensional face of the reduction body described uniquely by a subset of its inequality constraints treated as equalities.
    \item \textit{Face lattice}: The hierarchical arrangement of faces of the reduction body.
    \item \textit{Projection function} ($f_p$): A rank-deficient affine map from $\mathbb{Z}^{d} \rightarrow \mathbb{Z}^{d-a}$ defining to which element of the output each point in the reduction body accumulates.
    \item \textit{Accumulation space} ($\mathcal{A}$): The $a$-dimensional space characterized by the null space of the projection function.
    \item \textit{Dependence function} ($f_d$): An affine map from $\mathbb{Z}^{d} \rightarrow \mathbb{Z}^{d-r}$ characterizing from which element of the input each point in the reduction body reads.
    \item \textit{Reuse space} ($\mathcal{R}$): The $r$-dimensional space characterized by the null space of the dependence function.
    \item \textit{Reuse vector ($\rho$)}: Any vector in the reuse space.
\end{itemize}
Simplification is possible when the reuse space is non-empty (i.e., the dependence function, $f_{d}$, is rank-deficient).

\subsection{Face Lattice} \label{sec:background-face-lattice}

The face lattice~\cite{loechner_parameterized_1997} is an important data structure for simplification.
The face lattice of a polyhedron $\mathcal{D}$  is a graph whose nodes are the \emph{facets} of $\mathcal{D}$.
Each face in the lattice is the intersection of $\mathcal D$ with one or more \emph{equalities} of the form $\alpha z+\gamma = 0$ for $z \in \mathcal{D}$ obtained by \emph{saturating} one or more of the inequality constraints in $\mathcal{D}$.
We refer to each $k$-dimensional face as a \emph{($k$)-face}.

More than one constraint may be saturated to yield recursively, facets of facets, or \emph{faces}.
In the lattice, faces are arranged level by level, and each face saturates exactly one constraint in addition to those saturated by its immediate ancestors.

\subsubsection{Thick and Extended Faces}

GR06 introduces the notion of a thick face and what they call an \textit{effectively saturated constraint}.
An effectively saturated constraint is either a single equality constraint or a pair of parallel inequality constraints separated by a constant, which should be viewed as a single ``thick equality'' constraint.

For example, consider the following domains:
\begin{gather}
    \mathcal{D}_{1} = \{ [i,j] \mid (0 \leq i < N) \land (0 \leq j < N) \} 
    \label{eq:face-lattice-2d}
    \\
    \mathcal{D}_{2} = \{ [i,j] \mid (0 \leq i < 10) \land (0 \leq j < N) \}
    \label{eq:face-lattice-1d}
\end{gather}
The cardinalities of $\mathcal{D}_{1}$ and $\mathcal{D}_{2}$ are polynomials of the size parameter $N$.
There are $N^{2}$ points in $\mathcal{D}_{1}$ and there are $10N$ points in $\mathcal{D}_{2}$.
We say that $\mathcal{D}_{1}$ is 2-dimensional because the cardinality of $\mathcal{D}_{1}$ is quadratic in $N$.
Similarly, we view $\mathcal{D}_{2}$ as 1-dimensional because its cardinality is linear in $N$, and we say that the consraint $i=0$ is effectively saturated.
The dimensionality of an arbitrary polyhedron $\mathcal{D}$ is said to be its number of indices less its number of effectively saturated constraints.

\subsubsection{An Example}

\begin{figure}[tbh]
    \centering
    \includegraphics[width=0.47\textwidth]{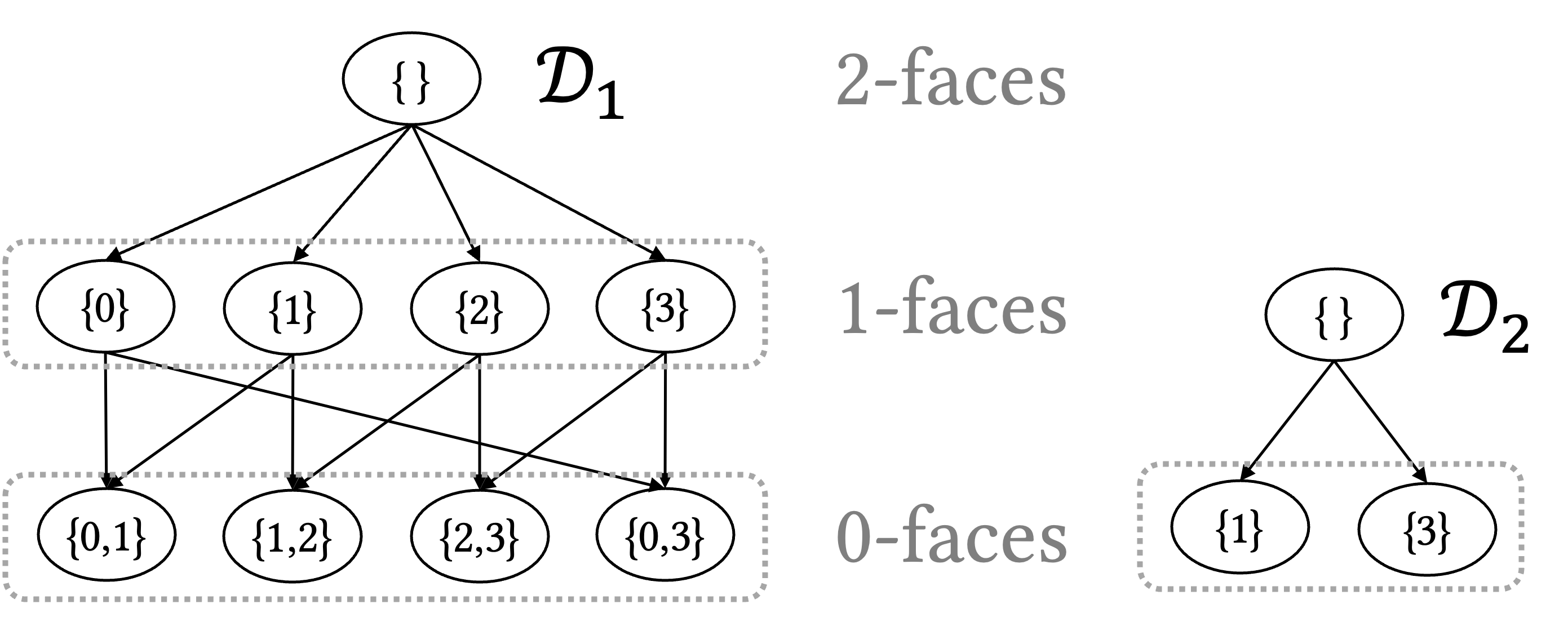}
    \caption{Face lattice of $\mathcal{D}_{1}$ in Equation~\ref{eq:face-lattice-2d}, the square with four edges (1-faces) and four vertices (0-faces) is shown on the left and $\mathcal{D}_{2}$ in Equation~\ref{eq:face-lattice-1d}, the thick line segment with two vertices (0-faces) on the right.}
    \Description{face lattice}
    \label{fig:face-lattices}
\end{figure}

Our implementation of the face lattice captures this notion of thick faces and constructs the face lattices for $\mathcal{D}_{1}$ and $\mathcal{D}_{2}$ as shown on the left and right respectively of Figure~\ref{fig:face-lattices}.
The four constraints of $\mathcal{D}_{1}$, $0 \leq i$, $0 \leq j$, $i < N$, and $j < N$, are denoted as 0, 1, 2, and 3.
Its four edges (or 1-faces) are described by saturating each constraint.
For example, the node ``\{0\}'' denotes saturing the 0'th constraint, $0 \leq i$, and represents the edge at $i=0$.
Similarly, the node ``\{0,1\}'' represents saturating both constraints, $0 \leq i$ and $0 \leq j$, and therefore represents the vertex at $[0,0]$.

\subsection{Single-Step Simplification} \label{sec:background-single-step-simplification}

\begin{figure}[tbh]
    \centering
    \includegraphics[width=0.25\textwidth]{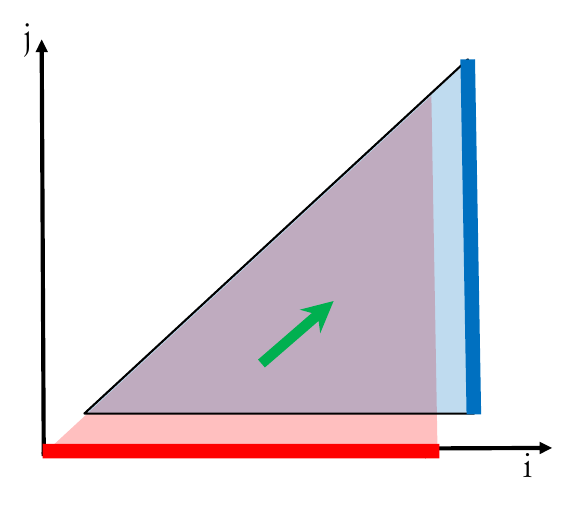}
    \caption{Simplification of a quadratic equation (computation defined over a triangle) to a linear complexity (the residual computation is defined only over the points in the bottom row).}
    \Description{Simplification of a quadratic equation (computation defined over a triangle) to a linear complexity (the residual computation is defined only over the points in the bottom row).}
    \label{fig:sr-example}
\end{figure}

Consider this example computing an $N$-element array $Y$:
\begin{equation}
  \label{eq:sr-example}
   Y_{i}=\sum_{j=0}^iX_{i-j}
\end{equation}
The equation has a reduction with the addition operator.
The reduction body is defined over the domain $\mathcal{D} = \{ [i,j] \mid 0\leq j\leq i<N \}$, the red triangle in Fig.~\ref{fig:sr-example}.

The $i$'th result, $Y_{i}$, is the accumulation of the values of the reduction body at points in the $i$'th diagonal of the triangle.
The complexity of the equation is the number of integer points in
$\mathcal{D}$, namely $O(N^2)$.
The reuse space is the null space of the indexing expression appearing inside the reduction body, which is $f_{d} : \{[i,j]\rightarrow [i-j]\}$ in this example.
Simplification is possible because the body has redundancy along the vector $\rho = [1, 1]$ (green arrow).
Any two points $[i, j], [i', j'] \in \mathcal{D}$ separated by a scalar multiple of $\rho$ read the same value of $X$ because $X_{i-j} = X_{i'-j'}$.
In other words, the body expression evaluates to the same value at all points along $[1, 1]$.  
Simplification exploits this reuse to read and compare only the $O(N)$ distinct values.
A geometric explanation of simplification is: 
\begin{itemize}
    \item  Translate $\mathcal{D}$ (shift the red triangle) by $\rho$ (green arrow) yielding $\mathcal{D}_s$ (the blue triangle).
    \item Delete all computations in the intersection of the two.
    \item Evaluate the residual computation on only the facets (here, edges) of $\mathcal{D}$ and/or $\mathcal{D}_s$.
\end{itemize}
Additionally, some of these facets can be ignored: the diagonal one, being parallel to $\rho$, was already included in the intersection, and the vertical (blue) one is \emph{external} since it does not contribute to any answer.
This leaves a residual computation on only the bottom facet.
Thus, we can replace the $O(N^2)$ equation by Equation \ref{eq:sr1}, which has only $O(N)$ complexity.
\begin{equation} \label{eq:sr1}
Y_i =
    \begin{cases}
        X_i & \text{if } i=0 \\
        Y_{i-1} + X_i & \text{if } i>0
    \end{cases}
\end{equation}
Indeed, Equation~\ref{eq:sr-example} specifies a simple scan (prefix-sum) of the input, albeit counter-intuitively.

\subsection{Equivalence Partitioning of Infinite Choices} \label{sec:background-equivalence-classes}

Notice that Equation~\ref{eq:sr1} expresses $Y_{i}$ in terms of $Y_{i-1}$.
It is equally possible to simplify Equation~\ref{eq:sr-example} by expressing $Y_{i}$ in terms of $Y_{i+1}$ instead.
There are two choices because the reuse space in this example is 1-dimensional (with two directions), and we could choose to exploit reuse along the vector $\rho_{1} = [1,1]$ or its negation, $\rho_{2} = [-1,-1]$.\footnote{Our compiler generates both.}  In the general case, when the reuse space is 2-dimensional or higher, the number of candidate choices is infinite.

However, not all choices of reuse need to be explored.
We illustrate this concretely in Section~\ref{sec:constructing-equivalence-classes}, but the intuition is that any two reuse vectors resulting in the same combination of residual computation can be viewed as the same candidate choice.  GR06 refers to the subset of $\rho_{i}$ that results in the same combination of residual computation as an \textit{equivalance class}.
Then, the set of all equivalence classes can be explored with dynamic programming.
As long as a single reuse vector from each equivalence class is considered, this is sufficient to guarantee the final simplified program's optimality.  GR06 does not give methods for constructing the set of equivalence classes.  Furthermore, the number of reuse vectors within a particular equivalence class is infinite, and we still must select one.  We address both these concerns in Section~\ref{sec:implementation}.

\subsection{Recursive Simplification} \label{sec:background-recursive-simplification}
In the general case, reductions have a $d$-dimensional reduction body, an $a$-dimensional accumulation, and an $r$-dimensional reuse space.
The process of Section \ref{sec:background-single-step-simplification} is applied recursively on the face lattice, starting with $\mathcal D$.
At each step, we simplify the facets of the current face $\mathcal F$.
The key idea is that exploiting reuse along $\rho$ avoids evaluating the reduction expression at most points in $\mathcal F$.
Specifically, let $\mathcal{F}'$ be the translation of $\mathcal F$ along $\rho$.
Then all the computation in $\mathcal{F}\cap \mathcal{F}'$ is avoided, and we only need to consider the two differences $\mathcal{F}'\backslash \mathcal{F}$ and $\mathcal{F}\backslash \mathcal{F}'$, i.e., the union of some of the facets of $\mathcal{F}$.  We are left with \emph{residual computations} defined only on (a subset of) the (thick/extended) facets of $\mathcal F$.

At each step of the recursion, the asymptotic complexity is reduced by exactly one polynomial degree, as facets of $\mathcal F$ are strictly smaller dimensional subspaces.
Furthermore, at each step, the newly chosen $\rho$ is linearly independent of the previously chosen ones.
Hence, the method is optimal---
all available reuse is fully exploited.
This holds regardless of the choice of $\rho$ at any level of the recursion, even though there may be infinitely many choices.

GR06 also develops several \emph{reuse exposing} transformations---e.g., reduction decomposition, exploiting distributivity---that enhance the opportunities to do simplification.  We implement all of them, as explained in Sec.~\ref{sec:case-studies}.

\section{Simplification in Practice} \label{sec:implementation}

In this section, we discuss methods for constructing the set of equivalence classes described in Section~\ref{sec:background-equivalence-classes} and selecting a candidate reuse vector from each.

\subsection{Constructing Equivalence Classes} \label{sec:constructing-equivalence-classes}

Simplification proceeds recursively down the face lattice, simplifying one facet at a time.
Let $\mathcal{F}$ be an arbitrary facet of the reduction body.
Let $\mathcal{F}_{i}$ denote the $i$'th facet of $\mathcal{F}$ (i.e., its $i$'th child), and let $\nu_{i}$ be the linear part of the normal vector of $\mathcal{F}_{i}$.
Let the symbol $\oplus$ be the reduction operator, and $\ominus$ be its inverse if $\oplus$ is invertible.  The single-step simplification of $\mathcal{F}$, summarized in Section~\ref{sec:background-single-step-simplification}, with the reuse vector $\rho$, results in a residual computation on some of its facets.
The orientation of $\rho$ relative to each facet $\mathcal{F}_{i}$ dictates the type of residual computation that occurs on $\mathcal{F}_{i}$.
There are three possibilities, depending on the sign of the dot product between $\rho$ and $\nu_{i}$:
\begin{enumerate}
    \item If $\rho \cdot \nu_{i} > 0$ then $\mathcal{F}_{i}$ contributes with the $\oplus$ operator.
    \item If $\rho \cdot \nu_{i} < 0$ then $\mathcal{F}_{i}$ contributes with the $\ominus$ operator.
    \item If $\rho \cdot \nu_{i} = 0$ then $\mathcal{F}_{i}$ does not contribute at all.
\end{enumerate}
Each facet $\mathcal{F}_{i}$ may be labeled as either an $\oplus$-face, $\ominus$-face, or $\oslash$-face respectively.
We say that $\rho$ \textit{induces a particular labeling}, $\mathcal{L}$, on the facets of $\mathcal{F}$.
Each labeling corresponds to one of the equivalence classes.
Given a particular labeling $\mathcal{L}$, we can construct the set of $\rho$ that induce $\mathcal{L}$ as the following:
\begin{equation} \label{eq:reuse-labeling}
    \mathcal{R}_{\mathcal{L}} = \mathcal{R} \cap \{ \rho \mid (\rho \cdot \nu_{i} > 0) \; \land \; (\rho \cdot \nu_{j} < 0) \; \land \; (\rho \cdot \nu_{k} = 0) \}
\end{equation}
where $\mathcal{R}$ is the overall reuse space and $\nu_{i}$ denotes the normal vectors of facets $\mathcal{F}_{i}$ labeled as $\oplus$-faces, $\nu_{j}$ as $\ominus$-faces, and $\nu_{k}$ as $\oslash$-faces.
In other words, we further constrain $\mathcal{R}$ to isolate the set of $\rho$ that induces the labeling $\mathcal{L}$.

Since each facet $\mathcal{F}_{i}$ of a face $\mathcal{F}$ can be given one of three labels, there are a total of $3^{m}$ labelings.  Our implementation currently enumerates all them, but there is a possibility for optimization (see Figure~\ref{fig:equivalence-classes-ex} which has only 12 possibilities rather than 27).  Some are impossible, e.g., there is no way to mark all facets with the same label.  

\subsubsection{An Example} \label{sec:3d-equiv-classes-ex}

\begin{figure}[tbh]
    \centering
    \includegraphics[width=0.47\textwidth]{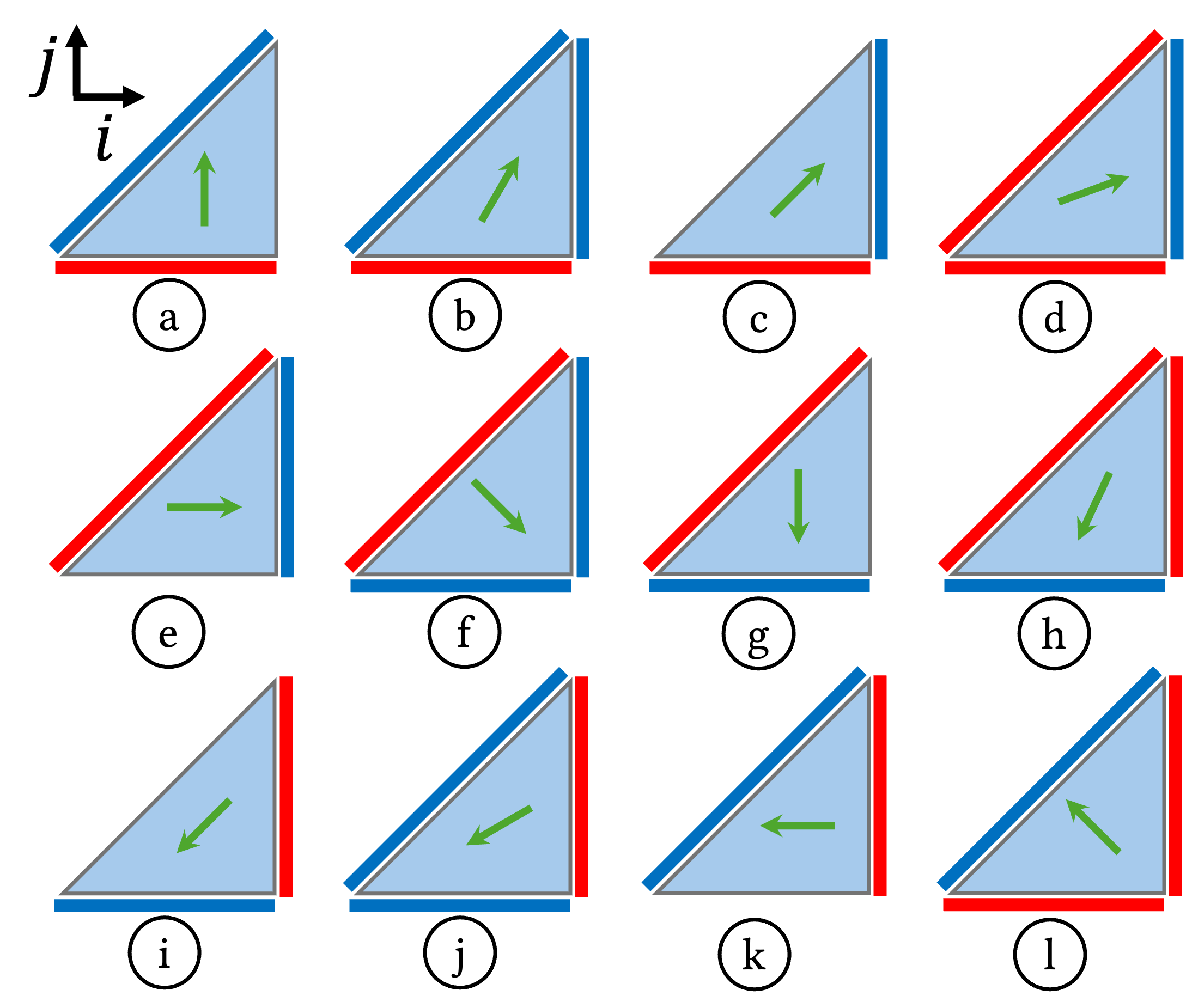}
    \caption{Equivalence classes explored for the single-step simplification of Equation~\ref{eq:reduction-equiv-classes}, with $\oplus$-faces highlighted in red, $\ominus$-faces in blue, and $\oslash$-faces not highlighted.}
    \Description{Equivalence classes}
    \label{fig:equivalence-classes-ex}
\end{figure}

Consider the 
\begin{equation} \label{eq:reduction-equiv-classes}
    Y_{i} = \sum_{(j,k)\geq 0}^{j\leq i \; \wedge \; k\leq i-j} X_{k}
\end{equation}
for $0 \leq i \leq N$.
The domain of the reduction body, $\mathcal{D}$, is:
\begin{equation} \label{eq:equiv-classes-domain-ex}
    \{ [i,j,k] \mid (0 \leq i \leq N) \land (0 \leq j) \land (k \leq i-j) \land (0 \leq k) \}
\end{equation}
The dependence function $f_{d}$ is 
    $\{ [i,j,k] \rightarrow [k] \}$,
and the reuse space,  $\mathcal{R}$ is the $ij$-plane: i.e., $\{ [i,j,k] \mid k=0 \}$, 

The tetrahedron $\mathcal{D}$ has four 2-faces.
The 2-face in the $ij$-plane (characterized by $k=0$) will always be labeled as a $\oslash$-face regardless of the choice of reuse, as any vector in the $ij$-plane is orthogonal to the normal vector of this face.
There are a total of 12 ways to label the remaining three 2-faces, illustrated in Figure~\ref{fig:equivalence-classes-ex}.

\subsection{Selecting Candidate Reuse Vectors}

Given a particular labeling, $\mathcal{L}$, the set of reuse vectors that induce the labeling, $\mathcal{R}_{\mathcal{L}}$, per Equation~\ref{eq:reuse-labeling} is either empty or unbounded.
From the perspective of a single-step simplification, there is no reason to prefer one $\rho \in \mathcal{R}_{\mathcal{L}}$ over another.
Any $\rho$ results in a decrease in asymptotic complexity by one polynomial degree, which follows from the optimality claim of GR06 
Therefore, we employ a simple heuristic that selects the smallest integer point closest to the origin.
Whether or not a $\rho$ can be chosen to guarantee that the constant factor is minimized remains an open problem.

\subsubsection{An Example}

The subsets of the reuse space that induce the 12 labelings illustrated in Figure~\ref{fig:equivalence-classes-ex} are given below, per the formulation in Equation~\ref{eq:reuse-labeling}:
\begin{align*}
\mathcal{R}_{a} &= \{ [i,j,k] \mid (k=0) \land (-i=0) \land (\textcolor{red}{j>0}) \land (\textcolor{blue}{i-j<0}) \} \\
\mathcal{R}_{b} &= \{ [i,j,k] \mid (k=0) \land (\textcolor{blue}{-i<0}) \land (\textcolor{red}{j>0}) \land (\textcolor{blue}{i-j<0}) \} \\
\mathcal{R}_{c} &= \{ [i,j,k] \mid (k=0) \land (\textcolor{blue}{-i<0}) \land (\textcolor{red}{j>0}) \land (i-j=0) \} \\
\mathcal{R}_{d} &= \{ [i,j,k] \mid (k=0) \land (\textcolor{blue}{-i<0}) \land (\textcolor{red}{j>0}) \land (\textcolor{red}{i-j>0}) \} \\
\mathcal{R}_{e} &= \{ [i,j,k] \mid (k=0) \land (\textcolor{blue}{-i<0}) \land (j=0) \land (\textcolor{red}{i-j>0}) \} \\
\mathcal{R}_{f} &= \{ [i,j,k] \mid (k=0) \land (\textcolor{blue}{-i<0}) \land (\textcolor{blue}{j<0}) \land (\textcolor{red}{i-j>0}) \} \\
\mathcal{R}_{g} &= \{ [i,j,k] \mid (k=0) \land (-i=0) \land (\textcolor{blue}{j<0}) \land (\textcolor{red}{i-j>0}) \} \\
\mathcal{R}_{h} &= \{ [i,j,k] \mid (k=0) \land (\textcolor{red}{-i>0}) \land (\textcolor{blue}{j<0}) \land (\textcolor{red}{i-j>0}) \} \\
\mathcal{R}_{i} &= \{ [i,j,k] \mid (k=0) \land (\textcolor{red}{-i>0}) \land (\textcolor{blue}{j<0}) \land (i-j=0) \} \\
\mathcal{R}_{j} &= \{ [i,j,k] \mid (k=0) \land (\textcolor{red}{-i>0}) \land (\textcolor{blue}{j<0}) \land (\textcolor{blue}{i-j<0}) \} \\
\mathcal{R}_{k} &= \{ [i,j,k] \mid (k=0) \land (\textcolor{red}{-i>0}) \land (j=0) \land (\textcolor{blue}{i-j<0}) \} \\
\mathcal{R}_{l} &= \{ [i,j,k] \mid (k=0) \land (\textcolor{red}{-i>0}) \land (\textcolor{red}{j>0}) \land (\textcolor{blue}{i-j<0}) \} \\
\end{align*}
The second piece, ``$(-i...)$'', in each of these comes the linear part of the constraint $N-i \geq 0$ from Equation~\ref{eq:equiv-classes-domain-ex}.
Similarly, the ``$(i-j...)$'' pieces should really be understood as $i-j-k...$ from Equation~\ref{eq:equiv-classes-domain-ex}, but we drop $k$ to save space since we also have the constraint $k=0$.
The constraints are colored to correspond with the edge colorings in Figure~\ref{fig:equivalence-classes-ex}.
Our heuristic selection criteria will choose $\rho_{a} = [0,1,0]$ from the first one, $\rho_{b} = [1,2,0]$ from the second one, etc., and  
 $\rho_{l} = [-1,1,0]$ for the last since these are the smallest integer points closest to the origin.

\subsection{Implementation Details}

\begin{figure}[tbh]
    \centering
    \includegraphics[width=0.47\textwidth]{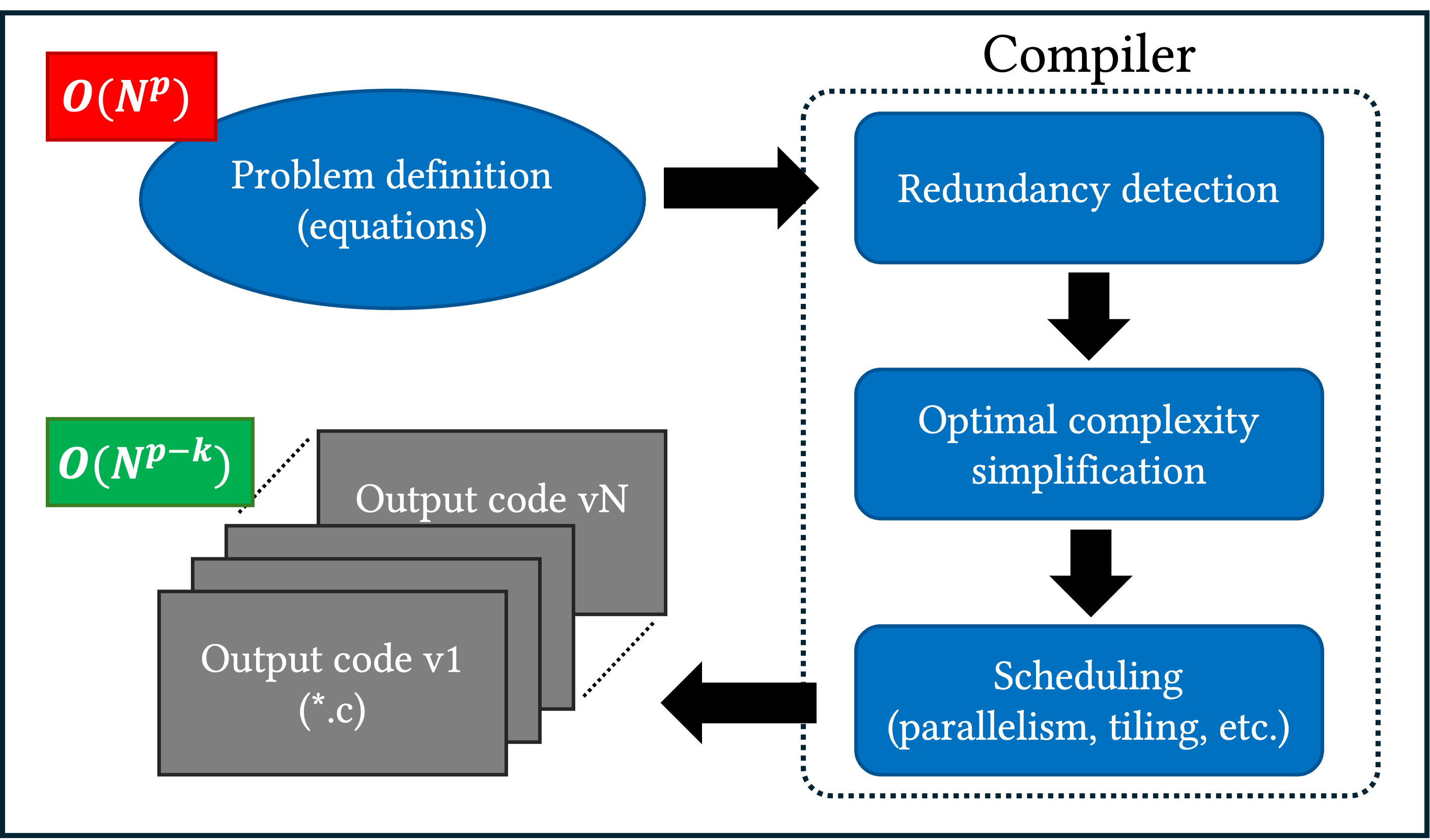}
    \caption{Alpha to C compilation pipeline.}
    \Description{Given an input equational program specification of complexity $O(N^{p})$, our compiler detects and eliminates any existing $k$-dimensional redundancy to produce a series (because there may be multiple) of \textit{redundancy-free} versions (v1, v2, etc.) each of optimal complexity $O(N^{p-k})$.}
    \label{fig:toolchain-flow}
\end{figure}

We implemented the simplification algorithm in the Alpha language~\cite{mauras_alpha_1989, le_verge_alpha_1991} and the AlphaZ system~\cite{yuki_alphaz_2013}, which supports reduction operations as first-class objects.
This provides concrete representations of the reduction body, projection function, dependence function, and face lattice described in Section~\ref{sec:background} using \texttt{isl} (the Integer Set Library~\cite{verdoolaege_isl_2010}) objects.
The semantics of an Alpha program closely follows the program's equivalent equational representation.
The workflow of our tool-chain is illustrated in Figure~\ref{fig:toolchain-flow}.
Given an input equational program specification, our compiler runs the GR06 algorithm and produces a series of transformed programs, one for each simplification possible.
We then generate C code for each transformed program to be compiled and run.

Consistent with GR06, we assume that the input program admits an equational representation (i.e., can be specified in the Alpha language) and that the program involves reduction operations.  An imperative polyhedral loop prgram can be easily brought into such a form (e.g., see the pioneering work of Feautrier~\cite{feautrier_dataflow_1991, redon_scheduling_1994, redon_detection_2000} on how this could be done).
We also assume that programs involve a single size parameter.

\section{Case Studies} \label{sec:case-studies}

In this section, we evaluate the efficacy of our implementation based on the design choices described in Section \ref{sec:implementation}.
We confirm that we can find simplified versions of all examples and generate code that produces the correct answers and exhibits the expected reduced asymptotic complexity.

Table \ref{table:simplification-summary} shows the initial runtime complexity and the expected complexity after simplification for each tested program.
We have created a specification for each example and confirmed that our compiler produces the expected simplifications.
We provide the original and simplified specifications with their corresponding generated C code in an accompanying artifact.

In all of the examples introduced below, we illustrate the algebra required to carry each aspect of simplification.
We do this only to provide the intution of how each step works.
We emphasize that all aspects below are completely automated as part of the implementation of the algorithm, and thus require no ``user input'' to guide the exploration.

\begin{table} [tbh]
    \centering
    \caption{Program simplification starting and expected asymptotic runtime complexities}
    \Description{Table indicating that all programs in the case study have a starting complexity of $O(N^3)$. The recursive simplification example has an expected final complexity of $O(N)$, and all others have a final complexity of $O(N^2)$.}
    \label{table:simplification-summary}
    \begin{tabular}{ccc}
        \toprule
        \makecell{Program} &
        \makecell{Starting\\Complexity} &
        \makecell{Final\\Complexity} \\
        \midrule
        Recursive Simplification   & $O(N^3)$ & $O(N)$   \\
        Reduction Decomposition    & $O(N^3)$ & $O(N^2)$ \\
        Distributivity             & $O(N^3)$ & $O(N^2)$ \\
        Matrix Multiplication ABFT & $O(N^3)$ & $O(N^2)$ \\
        \bottomrule
    \end{tabular}
\end{table}

\subsection{Recursive Simplification}\label{sec:eval-recursive-sr}

Programs with multiple dimensions of reuse require recursive simplification (see Section \ref{sec:background-recursive-simplification}) to produce an optimal specification.
To test our implementation of this, consider the example from Section~\ref{sec:3d-equiv-classes-ex}, given again in Equation \ref{eq:double-scan-original}.
\begin{equation}\label{eq:double-scan-original}
\begin{gathered}
    Y_i = \sum_{\{j,k\} \in \mathcal{D}} A_k \\
    \mathcal{D} = \{[j,k] \mid
        (0 \leq j \leq i) \land
        (0 \leq k \leq i-j) \}
\end{gathered}
\end{equation}
A naive execution of this specification has a time complexity of $O(N^3)$.
However, this equation specifies a prefix scan of a prefix scan.
This can be performed in linear time by performing a prefix scan over our inputs, saving these results in a temporary variable, then performing a prefix scan over the temporary variable.
This is demonstrated by Equations \ref{eq:double-scan-simplified1} and \ref{eq:double-scan-simplified2}:
\begin{align}
    Y_i &= \begin{cases}
        Z_0 & \text{if } i=0 \\
        Z_i + Y_{i-1} & \text{if } i>0 \\
    \end{cases}
    \label{eq:double-scan-simplified1}
    \\
    Z_i &= \begin{cases}
        A_0 & \text{if } i=0 \\
        A_i + Z_{i-1} & \text{if } i>0 \\
    \end{cases}
    \label{eq:double-scan-simplified2}
\end{align}

Each application of simplification is only able to lower the asymptotic time complexity by one degree.
Thus, to get our time complexity down from $O(N^3)$ to $O(N)$, we will need to apply simplification at least twice via recursive simplification (see Section \ref{sec:background-recursive-simplification}).
Thus, our compiler being able to produce any correct linear-time programs verifies a correct implementation of this feature.

When given the specification for Equation \ref{eq:double-scan-original}, it is able to generate 16 programs with a linear runtime, including the version described by Equations \ref{eq:double-scan-simplified1} and \ref{eq:double-scan-simplified2}.
Thus, we can confirm that recursive simplification has been successfully implemented.

\subsection{Reduction Decomposition}\label{sec:eval-decomposition}

Recall from Section~\ref{sec:background}, the accumulation space of a reduction may be multidimensional.
In mathematical notation, such reductions are often written as directly nested reductions with the same operator (e.g., a sum of a sum).
Such reductions may require \textit{decomposition} to expose reuse and allow simplification to take place.
In short, reduction decomposition can be thought of as a change of basis over one or more of the reduction variables, followed by separating a multi-variable reduction into nested single-variable reductions.

Consider the example of Equation \ref{eq:double-max-original}:
\begin{equation}
\begin{gathered}\label{eq:double-max-original}
    Y_i = \max_{\{j,k\} \in \mathcal{D}} A_{j,k} \\
    \mathcal{D} = \left\{ [j,k] \mid
        (i \leq j \leq 2i) \land
        (i \leq k \leq 3i-j)
    \right\}
\end{gathered}
\end{equation}
While there is reuse along the $i$ dimension (since we're reading $A_{j,k}$), we are not able to perform simplification.
All possible labelings of the face lattice involve at least one necessary $\ominus$-face, meaning the transformation requires the nonexistent inverse of the $\max$ operation.

However, we can perform reduction decomposition to expose reuse that can lead to simplification.
This is done by performing a change of basis to introduce $m=j+k$ and replace all instances of $j$ accordingly.
We then rewrite this single 2-dimensional reduction into two 1-dimensional reductions such that the outer reduction is over $m$ and the inner is over $k$.
Performing this transformation, then isolating the inner reduction to its own variable, produces Equation \ref{eq:decomposition-1}:
\begin{align}\label{eq:decomposition-1}
    Y_i = \max_{m=2i}^{3i} Z_{i,m}
    &&
    Z_{i,m} = \max_{k=i}^{m-i} A_{m-k,k}
\end{align}

The equation for $Z$ can now be simplified where $Z_{i,m}$ is computed from $Z_{i+1,m}$, producing an $O(N^2)$ program per Equation \ref{eq:decomposition-simplified}:
\begin{equation}\label{eq:decomposition-simplified}
    Z_{i,m} = \max \bigl( Z_{i+1,m},\ A_{m-i,i},\ A_{i,m-i} \bigr)
\end{equation}
This is the only such program, and our compiler is able to find it automatically.
The program specification was tested and confirmed to compute the correct answers.
Thus, we can confirm the correct implemenation of reduction decomposition to enable simplification.

\subsection{Distributivity}\label{sec:eval-distributivity}

Consider the case where a reduction's body contains a binary operation that distributes over the reduction operator (e.g., multiplication inside a summation).
If one of the terms is invariant at all points within the reduction body, it can be factored out.
This in turn may expand the reuse space of the reduction, enabling simplification.

Equation \ref{eq:sum-of-product} below, which performs a summation over the product of two terms, demonstrates such a situation:
\begin{equation}\label{eq:sum-of-product}
\begin{gathered}
    Y_i = \sum_{\{j,k\} \in \mathcal{D}} \left( A_{i,j+k} \times B_{k,j} \right) \\
    \mathcal{D} = \{ [j,k] \mid
        (0 \leq j \leq i) \land (0 \leq k \leq i)
    \}
\end{gathered}
\end{equation}
When computing a particular $Y_i$, all points in the reduction where $j+k$ are the same will read the same value of $A$.
With the summation as written, we cannot exploit this.
However, we can apply the same reduction decomposition transformation as in Section \ref{sec:eval-decomposition} to help us expose the reuse.
Performing the change of basis to introduce $m=j+k$ and replace $j$ such that the outer sum is over $m$ results in Equation \ref{eq:distributivity-decomposed}:
\begin{equation}\label{eq:distributivity-decomposed}
    Y_i = \sum_{m=0}^{2i} \left( \sum_{k=\max(0,m-i)}^{\min(i,m)} \left( A_{i,m} \times B_{m-k,k} \right) \right)
\end{equation}

Now, notice that $A_{i,m}$ is invariant at all points of the inner summation over $k$.
We can exploit the fact that multiplication distributes over addition to pull this term out, then isolate the inner reduction to its own variable, producing Equation~\ref{eq:distributivity-simplified}:
\begin{align}\label{eq:distributivity-simplified}
    Y_i = \sum_{m=0}^{2i} A_{i,m} \times Z_{i,m}
    &&
    Z_{i,m} = \sum_{k=\max(0,m-i)}^{\min(i,m)} B_{m-k,k}
\end{align}

Simplification can be applied to $Z_{i,m}$, allowing each value to be computed in constant time.
There are two choices for how to reuse answers: either $Z_{i+1,m}$ or $Z_{i-1,m}$.
Our compiler is able to produce both of these programs automatically, and both specifications were found to compute the correct answers.
Thus, we can confirm that our compiler can correctly exploit the distributive property to enable simplification.

\subsection{ABFT Checksum for Matrix Multiplication}\label{sec:abft-derivation}
In this section, we explain how the checksum computation from Section~\ref{sec:motivating-examples-abft} is discoverable by simplification.
Let us start by defining the row checksums, $\gamma_{i}$, over the output, $C$, as the following:
\begin{equation} \label{eq:gamma-checksum}
    \gamma_{i} = \sum_{j=0}^{N-1} C_{i,j}
\end{equation}
Substituting $C_{i,j}$ with its definition from Equation~\ref{eq:mm} lets us express $\gamma_{i}$ as:
\begin{equation} \label{eq:mm-simp-target}
\begin{gathered}
    \gamma_{i} = \sum_{\{j,k\} \in \mathcal{D}} \big( A_{i,k} \times B_{k,j} \big) \\
    \mathcal{D} = \{ [j,k] \mid (0 \leq j<N) \land (0 \leq k<N) \}
\end{gathered}
\end{equation}
As written, the computation specified by Equation~\ref{eq:mm-simp-target} is cubic, $O(N^{3})$, but it can be systematically rewritten with quadratic, $O(N^{2})$, complexity.

Simplification discovers the following three things about Equation~\ref{eq:mm-simp-target}.
First, values are accumulated over a 2D space (over $j$ and $k$).
Second, the multiplication operator inside the reduction body distributes over the reduction addition operator.
Third and finally, the subexpression $A_{i,k}$ evaluates to the same value along one of the accumulation dimensions (i.e., $A_{i,k}$ is independent of $j$).
Putting everything together allows the problem to be decomposed into a reduction of reductions, such that the outer reduction accumulates over $k$ and the inner reduction over $j$.
\begin{equation}
\gamma_{i} = \sum_{k=0}^{N-1} \sum_{j=0}^{N-1} \big( A_{i,k} \times B_{k,j} \big)
\end{equation}
Then, the term $A_{i,k}$ may be factored out of the inner reduction, and the inner summation over $j$ can be separated into its own equation.
\begin{align}
    \gamma_{i} = \sum_{k=0}^{N-1} \big( A_{i,k} \times X_{k} \big)
    &&
    X_{k} = \sum_{j=0}^{N-1} B_{k,j}
\end{align}
Finally, these equations together give the $O(N^{2})$ complexity version of the same computation specified by Equation~\ref{eq:gamma-checksum}.
The fact that summing rows of $B$ is required to detect errors in a particular row of the output $C$ falls out systematically from the simplification of Equation~\ref{eq:mm-simp-target}.

Complementarily, we would want to perform a similar simplification of checksums over \textit{columns} of $C$ to pinpoint the error's location and fix it in the output.
We don't show it here because it is analogous to the simplification described above.
In the end, we end up with the original $O(N^{3})$
matrix product and four $O(N^{2})$ checksums (two pairs of checksums over the columns and rows of the inputs and outputs).
Note that this is precisely what Huang and Abraham~\cite{huang_algorithm-based_1984} propose, though we only illustrate the simplification of row checksums here because the problems are symmetric.
Complete and working code can be found in the accompanying artifact.



\section{Evaluation}\label{sec:eval-rna}

In this section, we will evaluate our compiler's ability to automatically (i) reproduce the \Lyngso{} \textit{fast-i-loops} algorithm, (ii) derive some totally unknown cubic algorithms, and (iii) validate the result experimentally.
To do this, we will first show how the improved algorithm can be derived by hand.
Then, we will briefly discuss our compiler's rediscovery of this algorithm, plus three alternative algorithms produced from the same specification.
Finally, we will empirically confirm that the simplified programs produce the correct results and that they exhibit the expected asymptotic runtime characteristics.

\subsection{Reproducing the Fast-i-Loops Algorithm}\label{sec:reproducing-fast-i-loops}

Recall the $O(N^4)$ Equation \ref{eq:rna-vbi-n4} for computing the energy of an interior loop structure (used in RNA secondary structure prediction) from Section \ref{sec:motivating-examples-1999-lyngso}, repeated here for convenience.
\begin{equation}\label{eq:rna-vbi-n4-copy}
    Y_{i,j} =
    \min_{i<p<q<j}\bigl(
        A_{p,q} +
        B_{p-i+j-q} +
        C_{|p-i-j+q|}
    \bigr)
\end{equation}

First, we can decompose this reduction by introducing $k=p-i+j-q$ (the indexing expression for $B$), replacing $p$ accordingly.
We let the outer minimization be over $k$ and the inner be over $q$.
\begin{equation}
    Y_{i,j} = \min_{\substack{2 \leq k < j-i}} \;\Biggl(
        \min_{j-k<q<j} \bigl(
            A_{i-j+k+q,q} +
            B_{k} +
            C_{|k-2j+2q|}
        \bigr)
    \Biggr)
\end{equation}

Notice that the $B_k$ term is now invariant within the inner minimization and thus can be factored out.
We can then isolate the inner minimization to its own variable.
\begin{gather}
    Y_{i,j} = \min_{2 \leq k < j-i} \bigl( B_{k} + Z_{i,j,k} \bigr)
    \\
    Z_{i,j,k} = \min_{j-k < q < j} \bigl( A_{i-j+k+q,q} + C_{|k-2j+2q|} \bigr)
\end{gather}

Finally, we can notice that $Z_{i+1,j-1,k-2}$ minimizes a subset of the terms that $Z_{i,j,k}$ does.
We can then apply simplification to rewrite $Z_{i,j,k}$ as the minimum of $Z_{i+1,j-1,k-2}$ and a constant number of additional points in the reduction body.
This gives us our desired $O(N^3)$ algorithm, repeated below.
\begin{equation}
    Z_{i,j,k} = \min \begin{pmatrix}
        A_{i+1, j-k+1} + C_{|-k+2|} \\
        A_{i+k-1,j-1} + C_{|k-2|} \\
        Z_{i+1,j-1,k-2}
    \end{pmatrix}
\end{equation}

\subsection{Newly Discovered Algorithms} \label{sec:eval-new-algos}
We developed a specification for the full $O(N^4)$ fast-i-loops algorithm, matching the equation format and variable names presented by Jacob et al.~\cite{jacob_rapid_2010}.
When given to our compiler, it was able to automatically discover this same algorithm that \Lyngso{} et al. described, plus three alternative algorithms which were previously unknown.
The specification and the code to derive all four simplified algorithms can be found in the accompanying artifact.

One of the newly discovered algorithms performs a different decomposition using the expression used to index $C$ instead of $B$.
That is, it introduced $l = p-i+q-j$.
Since this value may be negative (as $q<j$), and since its absolute value is used to access $C$, different answers must be reused if $l$ is positive or negative.
Equations \ref{eq:rna-alternative-y} and \ref{eq:rna-alternative-z} below are representative of the algorithm produced by our compiler.
We refer to this algorithm as the ``New Algorithm''.
\begin{equation}\label{eq:rna-alternative-y}
    Y_{i,j} = \min_{i-j+3 \leq l \leq j-i-3} \bigl( C_{|l|} + Z'_{i,j,l} \bigr)
\end{equation}
\begin{equation}\label{eq:rna-alternative-z}
    Z'_{i,j,l} = \begin{cases}
        \min \begin{pmatrix}
            Z'_{i-1,j-1,l+2} \\
            A_{i+l+1, j-1} + B_{l+2}
        \end{pmatrix}
        & \text{if } l \ge 0 \\[6mm]
        \min \begin{pmatrix}
            Z'_{i+1,j+1,l-2} \\
            A_{i+1, j+l-1} + B_{-l+2}
        \end{pmatrix}
        & \text{if } l < 0
    \end{cases}
\end{equation}

The final two algorithms are a hybrid approach.
In short, they split the minimization of $Y_{i,j}$ into two cases early on, based on the positive and negative values of $l = p-i+q-j$.
One of the splits use the \Lyngso{} simplification described in Section \ref{sec:reproducing-fast-i-loops}, while the other split uses the alternative simplification of Equations \ref{eq:rna-alternative-y} and \ref{eq:rna-alternative-z}.
These two algorithms are referred to as ``Hybrid 1'' and ``Hybrid 2''.

Our compiler produces polynomials for the total number of loop iterations performed by each program as a metric for estimating the performance of each algorithm.
These are listed in Table \ref{table:loop-iterations}.

\begin{table}
\centering
\caption{Polynomials for the total number of loop iterations in each RNA minimum free energy algorithm.}
\label{table:loop-iterations}
\begin{tabular}{cc}
\toprule
Program & Loop Iterations \\
\midrule
Original Algorithm & $\frac{1}{24} N^4 + \frac{1}{12} N^3 - \frac{1}{24} N^2 + \frac{35}{12} N - 1 $\\[2mm]
\Lyngso{} Algorithm & $N^3 - 5N^2 + 17N - 30$ \\[2mm]
New Algorithm   & $\frac{4}{3} N^3 - 8N^2 + \frac{62}{3} N - 14$ \\[2mm]
Hybrid 1 & $\frac{13}{12} N^3 - \frac{43}{8} N^2 + \frac{271}{24}N + \frac{1}{4} \lfloor \frac{N}{2} \rfloor - 3$ \\[2mm]
Hybrid 2  & $\frac{5}{4} N^3 - \frac{61}{8} N^2 + \frac{211}{8} N - \frac{1}{4} \lfloor \frac{N}{2} \rfloor - 41$ \\[2mm]
\bottomrule
\end{tabular}
\end{table}

\subsection{Empirical Verification of Expected Complexities}

Remember that an inefficient implementation of quicksort gives an unbounded speedup over the most highly optimized implementation of bubble sort.  It is easy to forget this fundamental fact about asymptotic complexity, especially in systems-oriented venues/publications where most papers achieve a speedup over prior work that is only a \emph{constant factor} (often, the factor is a few percentage points).  Experimentally validating such speedups is (rightfully) extremely and critically important.
 
GR06 presents the theory of how to transform a polyhedral program (we view program, algorithm, and equations synonymously) whose initial asymptotic complexity is $O(N^d)$ into an equivalent program whose asymptotic complexity is  $O(N^{d-k})$.
So, the only reasonable evaluation of our tool is to first  confirm that the transformed program is correct---produces the same answers, and then show that input and output programs have the claimed asymptotic complexity.
To do this, we produced single threaded, demand-driven C code~\cite{wilde_naive_1995} for the original $O(N^4)$ algorithm and all four $O(N^3)$ algorithms.  We compiled with GCC and measured the execution time.


Correctness of the results was determined by using the program for the original specification as the source of truth.  Random inputs were generated and given to both the original and simplified programs. The outputs between the programs were then checked for equality (within floating point precision). All of the simplified programs were found to produce the correct results for a variety of problem sizes over multiple executions.


All programs were executed on a Linux system equipped with an Intel Core i7-12700K CPU.
The amount of memory used during each execution was confirmed to be less than the amount of memory available via the Linux \texttt{time} utility\footnote{\url{https://www.man7.org/linux/man-pages/man1/time.1.html}}.
This ensured that memory paging did not affect performance.

\begin{figure}
\begin{tikzpicture}
\begin{loglogaxis}[
    xlabel=Problem Size ($N$),
    ylabel=Runtime (seconds),
    legend style={at={(0.45,-0.2)},anchor=north},
    legend columns=3,
]
    \addplot [mark=, dashed] table [x=N, y=N3, col sep=comma] {data/AverageRuntimes.csv};
    \addplot [mark=, dashed] table [x=N, y=N4, col sep=comma] {data/AverageRuntimes.csv};
    \addplot table [x=N, y=Original, col sep=comma] {data/AverageRuntimes.csv};
    \addplot table [x=N, y=Lyngso, col sep=comma] {data/AverageRuntimes.csv};
    \addplot table [x=N, y=New, col sep=comma] {data/AverageRuntimes.csv};
    \addplot table [x=N, y=Hybrid1, col sep=comma] {data/AverageRuntimes.csv};
    \addplot table [x=N, y=Hybrid2, col sep=comma] {data/AverageRuntimes.csv};

    \legend{
        $N^3$,
        $N^4$,
        Original,
        \Lyngso,
        New Algorithm,
        Hybrid 1,
        Hybrid 2}
\end{loglogaxis}
\end{tikzpicture}
\caption{Average runtimes of all generated fast-i-loops programs, plotted alongside $(\expnumber{7}{-9})N^3$ and $(\expnumber{1}{-10})N^4$ to verify the polynomial degree of the runtimes with respect to the problem size $N$.}
\Description{Average runtimes of all generated  fast-i-loops programs, plotted alongside $(\expnumber{7}{-9})N^3$ and $(\expnumber{1}{-10})N^4$ to verify the polynomial degree of the runtimes with respect to the problem size $N$.}
\label{fig:runtimes}
\end{figure}
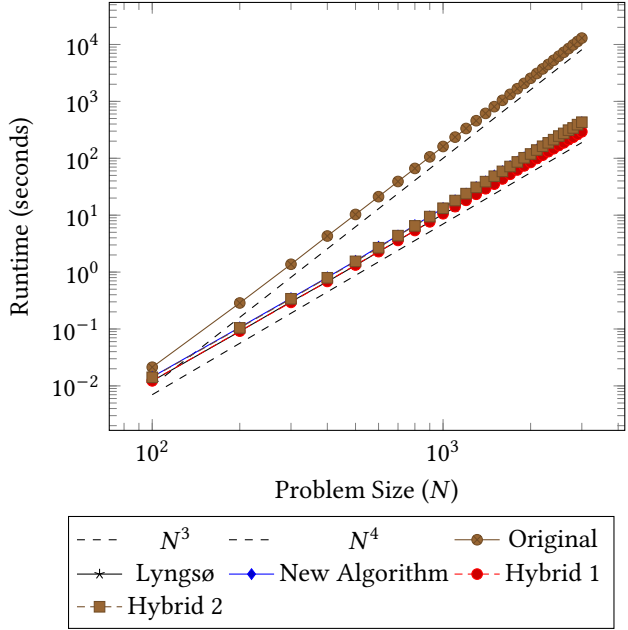

Figure \ref{fig:runtimes} shows a plot for the algorithm runtimes across problem sizes from $N=100$ to $N=3000$ in increments of $100$.
This plot uses logarithmic scaling for both axes, which causes polynomials to be shown as straight lines.
Functions for $O(N^3)$, and $O(N^4)$ are plotted alongside the results to visually confirm that the measured complexities match what we theoretically expected.
That is, the original specification runs in $O(N^4)$ time, while the simplified versions run in $O(N^3)$ time.

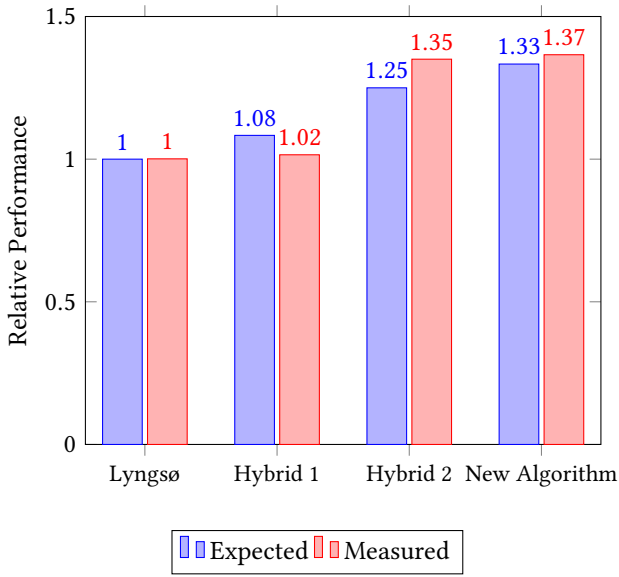
\begin{figure}
\begin{tikzpicture}
\begin{axis}[
ybar,
ymin=0,
ymax=1.5,
ylabel={Relative Performance},
symbolic x coords={\Lyngso{},Hybrid 1,Hybrid 2,New Algorithm},
xtick=data,
x tick label style = {font=\small},
nodes near coords,
nodes near coords align={vertical},
legend style={at={(0.45,-0.2)},anchor=north},
legend columns=2,
bar width=15pt,
enlarge x limits=0.15,
]
\addplot coordinates {
    (\Lyngso{},1)
    (Hybrid 1,1.08333333333333)
    (Hybrid 2,1.25)
    (New Algorithm,1.33333333333333)
};
\addplot coordinates {
    (\Lyngso{},1.00107176834485)
    (Hybrid 1,1.01529324775785)
    (Hybrid 2,1.3503718512292)
    (New Algorithm,1.36599328989149)
};
\legend{Expected,Measured}
\end{axis}
\end{tikzpicture}
\caption{Relative performance of the simplified fast-i-loops programs averaged across all problem sizes. }
\Description{Relative performance of the simplified fast-i-loops programs averaged across all problem sizes. Expected and measured values are computed per Table~\ref{table:loop-iterations} and the relative program runtime, respectively.}
\label{fig:runtime-ratios}
\end{figure}



Based on the polynomials for the number of loop iterations per program (see Table \ref{table:loop-iterations}), we expect the \Lyngso{}-equivalent program to have the best performance, followed closely by the ``Hybrid 1'' program.
The ``Hybrid 2'' and ``New Algorithm'' programs are expected to be about $25\%$ and $33\%$ slower respectively.
To evaluate the accuracy of this, Figure \ref{fig:runtime-ratios} reports the measured performance of each program relative to the fastest, next to the theoretically expected ratios.
The greatest error in this estimation was for the ``Hybrid 2'' algorithm, where our measured performance differed from the expected performance by $7.72\%$.
This shows that, for the tested programs, the execution time is an acceptable proxy for the total number of operations performed.



\subsection{Compile Times}

In general, there may be many different simplified versions of the programs, like the four cubic simplifications of the fast-i-loops in Table~\ref{table:loop-iterations}.
We report the time that it took our implementation to find the simplifications in Table~\ref{table:compile-times}.
The ``First'' column shows the time it took our implementation to find at least one valid simplification.
The ``All'' column reports the time to explore all possible paths.
Note that, in general, it may take prohibitively long to exhaustively search the space.
We leave considerations about navigating and pruning the search space as future work, as such our currently simply enumerates and explores all possible paths in this work.

\begin{table} [tbh]
    \centering
    \caption{All times measured in seconds. ``LoC'' denotes ``lines of code'' in the generated C code of the simplified programs.}
    \label{table:compile-times}
    \begin{tabular}{ccccc}
        \toprule
        \makecell{Section} &
        \makecell{First (sec)} &
        \makecell{All (sec)} &
        \makecell{Avg. LoC} &
        \makecell{gcc (sec)} \\
        \midrule
        \ref{sec:eval-recursive-sr}   & 8.1 & 19.6 & 254 & 0.21 \\
        \ref{sec:eval-decomposition}  & 4.9 & 5.3 & 184 & 0.34 \\
        \ref{sec:eval-distributivity} & 5.3 & 8.3 & 240 & 0.27 \\
        \ref{sec:eval-new-algos}      & 8.3 & 33.6 & 738 & 1.28 \\
        \bottomrule
    \end{tabular}
\end{table}

\section{Related Work} \label{sec:related-work}

Simplification has garnered renewed interest recently.  Asymptotic inefficiencies are present, even in deployed codes.  Ding and Shen~\cite{ding_glore_2017} noted that nine of the 30~benchmarks in Polybench~3.0, and two deployed PDE solvers have such inefficiencies. Separately, Yang, et al.~\cite{yang_simplifying_2021} showed that simplification is useful for many algorithms in statistical learning like Gibbs Sampling (GS), Metropolis Hasting (MH) and Likelihood Weighting (LW).  Their bencharks include Gaussian Mixture Models (GMM), Latent Dirichlet Allocation (LDA) and Dirichlet Multinomial Mixtures (DMM).  See their paper for details of benchmarks, algorithms, size parameters, machine specs, etc.



There is ongoing research into the RNA secondary structure algorithms.
The optimizations developed by \Lyngso{} et al.\ have been implemented by a variety of others \cite{dirks_partition_2003,fornace_unified_2020,jacob_rapid_2010}.
Even so, there are applications that have not yet incorporated this algorithmic improvement \cite{lorenz_viennarna_2011,mathews_using_2004}.
Separately, work is being done to schedule and optimize these and similar equations within the polyhedral model \cite{palkowski_parallel_2017,wonnacott_automatic_2015}.
However, there still exist other algorithms in this domain which do not have known optimized versions but could be optimized by our compiler, such as the Maximum Expected Accuracy (MEA) equations \cite{lu_improved_2009,palkowski_parallel_2020}.

Simplification is related but complementary to the problems of \emph{marginalization of product functions} (MPF)  and its discrete version,  \emph{tensor contraction} (TC).  Such problems arise in many domains.  MPF can be optimally computed using Pearl's ``summary passing'' or ``message passing'' algorithm~\cite{pearl_probabilistic_1988} for Bayesian inference, or the generalized distributive law~\cite{aji_generalized_2000, kschischang_factor_2001, shafer_probability_1990}.
Similarly, there is a long history of research on optimal implementations of TC~\cite{pfeifer_faster_2014,kong_automatic_2023}.  In this problem, the sizes of the tensors in each dimension are known, and we seek the implementation with the minimum number of operations.   

To explain the problem, first note that the cost of multiplying three matrices, $A$, $B$ and $C$ is affected by how associativity is exploited: if A and C are short-stout, and B is tall-thin, $(AB)C$ is better than $A(BC)$, and if A and C are tall-thin and B is short-stout, the latter parenthesization is.  Indeed, optimization of this for a sequence of matrices is a classic, textbook problem, used to illustrate dynamic programming.  But the underlying matrix multiplication uses the standard, cubic algorithm.  TC extends this to multiple chains of products (requiring  us to optimally identify and exploit common subexpressions) and to tensors, rather than matrices (exposing opportunities to exploit simplification by inserting new varibles).

Specifically, consider the following system of equations:
\begin{align}
    X_{i,l} = \sum_{j,k=1}^{N}A_{i,j,l} \times B_{i,k}
    &&
    Y_j = \sum_{i,k,j=1}^{N}A_{i,j,l} \times B_{i,k}
\end{align}
Naively, each equation would have $O(N^{4})$ complexity since each result is the accumulation of a triple summation, and there are  $O(N)$ answers. However, if we define two new variables:
\begin{align}
    T_{i,l}  = \sum_{j=1}^{N}A_{i,j,l}
    &&
    T'_{i} = \sum_{k=1}^{N}B_{i,k}
\end{align}
then the following is an equivalent simplified system of equations whose complexity is only $O(N^{3})$:
\begin{align}
    X_{i,l} = T_{i,l} \times T'_{i}
    &&
    Y_j = \sum_{i,l=1}^{N}A_{i,j,l} \times T'_{i}
\end{align}

\section{Future Work} \label{sec:future-work}

Our compiler works only on \emph{independent} reductions, i.e., those where the expressions/values over which the reduction operator is applied do not use \emph{any instance} of the result variable.  However, Yang et al.~\cite{yang_simplifying_2021} show that dealing with \emph{dependent} reductions, where some outputs may depend recursively on computed results, is an important problem in practice.  Our compiler currently may produce equations that do not admit a legal schedule. This can, of course, be detected by running a scheduler on the simplified program. We are working on integrating scheduling with simplification in the standard recursion down the face lattice.

Consistent with GR06, we only handle programs with a single size parameter.  
Extending the theory to multiple size parameters is important for many algorithms, e.g., in RNA computations, we may have two sequences of respective lengths, $N$ and $M$.  
As noted by Loechner and Wilde~\cite{loechner_parameterized_1997}, such domains may be decomposed into \textit{chambers}, each with a distinct face lattice, hence polyhedral tools are capable of handling multiple parameters.
However, the function describing complexity is still a multivariate polynomial involving two non-comparable parameters, $N$ and $M$.
This introduces a partial order among the complexity functions and requires extending the core simplification algorithm.

\section{Conclusion} \label{sec:conclusion}

Nearly two decades after the theory was first proposed by Gautam and Rajopadhye~\cite{gautam_simplifying_2006}, we implemented the reduction simplification transformation, a powerful program transformation capable of automatically exploiting reuse in programs involving reductions.  The original theory omitted several details required for employing simplification in practice, which we discussed and addressed, thus providing the first complete push-button implementation of simplification in a compiler.

We evaluated its effectiveness in automatically rediscovering several key results in algorithmic improvement previously only attainable through manual human analysis.  In particular, we illustrated how simplification discovered three new cubic algorithms for RNA secondary structure prediction.  We also showed how it can be used to fully automate the work on algorithm-based fault tolerance (ABFT) originally proposed by Huang and Abraham~\cite{huang_algorithm-based_1984} and subsequently employed by a community of authors for a range of problems~\cite{zhao_ft-cnn_2021, davies_correcting_2018, hakkarinen_fail-stop_2015, wu_towards_2016}.  Our work takes a step toward raising the level of abstraction for the user and demonstrates how optimizations that previously required clever, painstaking analysis can be systematically employed as a sequence of compiler transformations.


\balance
\bibliographystyle{ACM-Reference-Format}
\bibliography{zotero-ln}

\end{document}